# An Empirical Analysis of Google Play Data Safety Disclosures: A Consistency Study of Privacy Indicators in Mobile Gaming Apps

**Bakheet Aljedaani**[1]

[1] Computer Science Department, Jamoum University College, Umm Al-Qura University, Makkah, Saudi Arabia

Corresponding author: Bakheet Aljedaani (e-mail: bhjedaani@uqu.edu.sa).

**ABSTRACT** The Google Play marketplace has introduced the Data Safety section to improve transparency regarding how mobile applications (apps) collect, share, and protect user data. This mechanism requires developers to disclose privacy and security-related practices, including data collection, data sharing, and data protection measures. However, the reliability of these disclosures remains dependent on developer self-reporting, raising concerns about their accuracy. This study investigates the consistency between developer-reported Data Safety disclosures and observable privacy indicators extracted from Android Application Packages (APKs). An empirical analysis was conducted on a dataset of 41 mobile gaming apps, including 21 children-oriented and 20 general-audience apps. A static analysis approach was used to extract key privacy indicators from APK files, including device identifiers, data sharing practices, personal information access, and location access. These indicators were systematically compared with the corresponding disclosures reported in the Google Play Data Safety labels using a structured consistency evaluation framework. The results revealed varying levels of agreement across privacy categories. Device identifier disclosures demonstrated relatively high consistency (87.8%), whereas other indicators exhibited substantial mismatches. In particular, location-related disclosures showed the highest inconsistency rate (56.1%), followed by personal information and data sharing indicators. Comparative analysis between children-oriented and general-audience apps revealed similar mismatch patterns across both categories. Furthermore, Chi-square statistical tests indicate that these differences are not statistically significant, suggesting that disclosure inconsistencies are not associated with app category but instead reflect broader ecosystem-level challenges. These findings highlight limitations in the reliability of current marketplace transparency mechanisms and emphasize the need for improved validation and verification approaches to ensure accurate privacy reporting in mobile app ecosystems.

## I. INTRODUCTION

Mobile applications (apps) have become an integral part of modern digital ecosystems, supporting a wide range of services including healthcare, entertainment, education, and commerce. To enable these services, mobile apps frequently collect and process user data for purposes (e.g., analytics, advertising, personalization, and service improvement) [1], [2]. As a result, concerns regarding the transparency of data collection and sharing practices have become increasingly important. End-users often rely on information presented in app marketplaces to understand how their data may be handled before installing an app. However, determining whether such disclosures accurately reflect actual app behavior remains a challenging problem [3]. Prior research has shown that mobile apps often access sensitive resources and user data in ways that are not always clearly communicated to end-users, raising concerns about privacy awareness and informed consent [4], [5]. In addition, studies have highlighted that permission systems alone do not provide sufficient insight into how user data is actually processed or shared within apps [6], [7].

To address these concerns, Google introduced the Data Safety section in the Google Play marketplace, requiring developers to disclose how their apps collect, share, and protect user data [8]. This feature provides structured information about data practices, including whether apps collect sensitive data (e.g., location or personal information, whether data is shared with third parties, and whether it is encrypted during transmission). The introduction of Data Safety labels represents an important step toward improving transparency and enabling end-users to make more informed decisions when selecting mobile apps. However, recent study suggests that privacy labels may not fully address users' actual privacy concerns, and found that a significant portion of users' privacy questions remain unanswered or only partially addressed by existing mobile app privacy labels [9]. Similar transparency initiatives have been explored in prior work, including efforts to improve privacy notices, standardized disclosure formats, and machine-readable privacy representations [1], [10] and [11]. However, the effectiveness of such mechanisms depends on the accuracy and completeness of developer-reported disclosures, which may vary depending on developer interpretation and implementation of reporting requirements [12]. Despite these improvements, the reliability of developer-reported privacy disclosures remains uncertain. This challenge is further compounded by the difficulty developers face in accurately completing privacy labels, particularly when apps integrate complex third-party components and data processing mechanisms [13]. The information presented in Data Safety labels is self-reported and not automatically verified by the platform, raising concerns about potential inconsistencies between declared practices and actual app behavior. Recent work has demonstrated that may not fully reflect observable app behavior and found that a substantial number of mobile apps transmit data that is not declared in their privacy labels, highlighting the lack of validation in current marketplace transparency mechanisms [14]. Previous studies have demonstrated discrepancies between privacy policies and observable app characteristics, including permission usage, data flows, and third-party tracking behaviors [1], [15]. For example, studies have shown that many mobile apps request permissions that are not clearly justified by their functionality or fail to disclose certain data practices in their privacy policies [16], [17]. Other work has highlighted the role of third-party libraries and advertising frameworks in introducing additional data collection capabilities that may not be fully understood or disclosed by developers [15], [18] and [19]. These findings suggest that transparency mechanisms based solely on developer self-reporting may not always provide an accurate representation of app privacy behavior.

In this context, this study investigates the consistency between developer-reported Google Play Data Safety disclosures and observable privacy indicators extracted from Android Application Packages (APKs). Unlike prior work that focuses primarily on permissions or privacy policies, this study provides a direct comparison between marketplace disclosures and application-level technical indicators, enabling a more comprehensive assessment of privacy transparency. The analysis focuses on mobile gaming apps, one of the most widely adopted categories of mobile apps, particularly among children and young end-users [20], [21]. Given the scale and popularity of mobile games, as well as their increasing reliance on analytics and advertising infrastructures, understanding the reliability of their privacy disclosures is of significant importance. To provide deeper insights, the study includes both children-oriented and general-audience mobile gaming apps, enabling a comparative analysis of disclosure practices across app categories. In addition, the study applies statistical analysis techniques, including Chi-square tests, to evaluate whether observed differences in disclosure mismatches are statistically significant. By combining static analysis of APK files with structured analysis of marketplace disclosures, this study aims to provide empirical evidence regarding the effectiveness of Data Safety labels as a transparency mechanism in mobile app ecosystems. This study addresses the following research questions (RQs):

***RQ1:*** *To what extent do Google Play Data Safety disclosures align with observable static privacy indicators in mobile gaming apps?*

This RQ aims to measure the level of consistency between developer-reported privacy disclosures and indicators extracted through static analysis of APK files.

***RQ2:*** *Do disclosure inconsistencies differ between children-oriented and general-audience apps?*

This RQ aims to examine whether app category influences the likelihood of mismatches between static indicators and Data Safety labels.

***RQ3:*** *Are the observed differences in disclosure mismatches statistically significant?*

This RQ aims to apply statistical analysis techniques to determine whether disclosure inconsistencies are associated with app category or occur broadly across the dataset.

To answer these RQs, we conducted an empirical study combining static analysis of Android apps with structured analysis of Google Play Data Safety disclosures. The study was performed on a selected dataset of 41 widely installed Android mobile gaming apps, including both children-oriented and general-audience titles. The dataset was divided into two groups: 21 children-oriented apps and 20 general-audience apps, selected based on popularity indicators such as installation counts and marketplace rankings to ensure representation of widely used apps with large user bases. The study was carried out through a four-phase workflow, including two main technical parts: static analysis and Data Safety extraction. The static analysis was performed for APK files to extract privacy-related indicators, including device identifier usage, personal information access, data sharing signals, and location access, based on permissions and observable app artifacts. Then, we manually collected and analyzed the corresponding Google Play Data Safety disclosures, mapping developer-reported information into structured indicators aligned with the static analysis results. The extracted data were then systematically compared to identify mismatches between observable indicators and reported disclosures, followed by comparative analysis across app categories and statistical validation using Chi-square testing. The results reveal that while certain indicators, such as device identifiers, exhibit relatively high agreement, other categories including location, personal information, and data sharing show substantial inconsistencies between developer-reported disclosures and observable app characteristics. Furthermore, similar mismatch patterns were observed across both children-oriented and general apps, suggesting that privacy transparency challenges are not limited to specific app categories but may reflect broader issues in marketplace disclosure mechanisms. We believe this study makes the following key contributions:

- *Empirical comparison of privacy disclosures and technical indicators:* This work provides a systematic comparison between developer-reported Google Play Data Safety disclosures and observable static privacy indicators, enabling a direct assessment of the reliability of marketplace transparency mechanisms.
- *Cross-category analysis of mobile gaming apps:* The study analyzes both children-oriented and general-audience apps, providing insights into whether privacy transparency differs across app categories subject to different regulatory and design considerations.
- *Identification of systematic disclosure inconsistencies:* The results reveal consistent mismatch patterns across multiple privacy indicators, highlighting potential limitations in current self-reported disclosure mechanisms within mobile application marketplaces.
- *Integration of statistical analysis for validation:* The study applies Chi-square statistical testing to evaluate whether observed differences are statistically significant, providing a rigorous empirical basis for interpreting the findings.
- *Implications for improving privacy transparency mechanisms:* The findings provide actionable insights for developers, platform providers, and researchers, emphasizing the need for improved tooling, clearer reporting guidelines, and potential automated validation mechanisms.

The rest of this paper is organized as follows: Section II presents the most relevant works to our study. Section III details the adopted research methodology. Section IV reports the study findings, followed by a discussion in Section V. Threats to validity are discussed in Section VI, and Section VII concludes the paper, and suggests potential future work.

## II. RELATED WORK

Research on mobile app privacy and security has been widely explored from multiple perspectives, including permission analysis, data flow analysis, privacy policies, third-party tracking, and transparency mechanisms. This section reviews relevant studies and positions the current work within the existing literature.

### A. Permission-Based Privacy Analysis in Mobile Apps

Early research on mobile privacy primarily focused on analyzing Android permission systems as indicators of app behavior. Felt et al. in [4] conducted one of the earliest systematic studies of Android permissions, highlighting the gap between permission requests and user understanding. Similarly, Kelley et al. in [5] investigated end-user perception of permission requests, showing that end-users often struggle to interpret their implications. Subsequent studies demonstrated that permissions alone are insufficient to accurately characterize privacy risks, as apps may request unnecessary permissions or use them in unexpected ways [22], [7]. Further work extended permission-based analysis by examining how permissions relate to app functionality and privacy exposure. Zhou and Jiang in [17] analyzed Android malware and demonstrated how permissions can be abused for malicious purposes. Enck et al. in [23] introduced static analysis techniques to identify potential privacy leaks in Android apps. However, several studies have concluded that permission-based approaches do not represent privacy behavior with sufficient precision, as they do not capture how data is processed or shared after being accessed [24], [25].

### B. Static and Dynamic Analysis of Privacy Behavior

To overcome the limitations of permission-based approaches, researchers have developed static and dynamic analysis techniques to examine how apps handle end-user's data. Tools such as FlowDroid [16] and TaintDroid [23] enabled precise tracking of sensitive data flows within Android apps. These techniques have been widely used to detect privacy leaks and unauthorized data transmissions. Dynamic analysis approaches, such as those proposed by Rastogi et al. in [26] and Ren et al. in [27] have further demonstrated that many apps transmit sensitive data to remote servers, often without clear end-user awareness. While these approaches provide deeper insights into app behavior, they are often computationally expensive and difficult to scale across large datasets. In contrast, static analysis remains a widely used technique due to its scalability and ability to analyze large numbers of apps efficiently [28]. However, both approaches face challenges in capturing behaviors introduced by obfuscation, dynamic code loading, and third-party libraries.

### C. Third-Party Libraries and Tracking Ecosystems

A significant body of research has examined the role of third-party libraries, particularly advertising and analytics SDKs, in shaping privacy risks in mobile apps. Studies by Grace et al. in [29] and Wang et al. in [19] showed that third-party components frequently introduce additional data collection and tracking capabilities that are not always transparent to users. Similarly, Ikram et al. in [30] analyzed mobile advertising ecosystems and found widespread tracking behavior across apps. Recent studies have highlighted that developers may not fully understand the privacy implications of integrated SDKs, leading to incomplete or inaccurate disclosures of data practices [31], [32]. These findings are particularly relevant in the context of mobile games, where monetization strategies often rely heavily on advertising and analytics infrastructures. As a result, third-party libraries represent a key factor contributing to the complexity of privacy transparency in mobile ecosystems.

### D. Privacy Policies and Transparency Mechanisms

Another line of research has focused on evaluating privacy policies and user-facing transparency mechanisms. Zimmeck et al. in [10] developed automated techniques for analyzing privacy policies and identifying inconsistencies between stated and actual practices. Cranor in [11] discussed the challenges of designing effective privacy notices, emphasizing that traditional policy formats are often difficult for users to understand. More work has explored structured and machine-readable approaches to privacy disclosures, aiming to improve usability and transparency [33], [34]. However, even when structured disclosures are available, users may still struggle to interpret their meaning or assess their implications [35]. These challenges highlight the importance of evaluating not only the presence of privacy disclosures but also their accuracy and reliability.

### E. Studies on Google Play Data Safety Labels

With the introduction of Google Play Data Safety labels, research has begun to examine their effectiveness as a transparency mechanism. While this feature is relatively recent, emerging studies suggest that privacy labels may face important limitations in both accuracy and usability. For instance, Koch et al. in [14] conducted a large-scale empirical study analyzing the consistency between Apple's privacy labels and actual app behavior through network traffic analysis. Their results showed that a significant number of apps transmit data not declared in their privacy labels, indicating discrepancies between developer-reported disclosures and observable behavior. The study also highlighted that privacy labels are not systematically validated during the app review process, raising concerns about their reliability as a transparency mechanism. Zhang and Sadeh in [9] evaluated the effectiveness of mobile app privacy labels in addressing users' privacy questions through an empirical analysis of a large corpus of user concerns. Their findings indicate that existing privacy labels often fail to provide sufficient information to answer many user questions, with a substantial portion of questions either unanswered or only partially addressed. These results suggest that, beyond issues of accuracy, privacy labels may also suffer from limitations in completeness and usability.

Cross-platform analyses have also highlighted inconsistencies in privacy label disclosures across different mobile ecosystems. Khandelwal et al. in [36] compared privacy labels of apps available on both Android and iOS platforms and found notable differences in how data practices are reported across marketplaces. Their findings suggest that privacy disclosures may vary depending on platform-specific guidelines and developer interpretations, further raising concerns about the consistency and reliability of privacy labels. Lin et al. in [12] conducted a user-centered study comparing Android Data Safety labels with iOS privacy labels and found that, although some end-users perceive the labels as helpful, many users find them vague, incomplete, and difficult to interpret, leading to misunderstandings of app data practices. The study also reported that end-users often miss critical information due to interface design and express uncertainty regarding the reliability of developer-provided disclosures.

Furthermore, Zimmeck et al. in [10] and Reidenberg et al. in [35] demonstrated that privacy disclosures and policies often fail to accurately reflect actual data practices, highlighting limitations in user-facing transparency frameworks. Similarly, empirical analyses of mobile apps have shown that developers may over-report or under-report data collection practices, partly due to ambiguity in reporting requirements and limited visibility into third-party library behavior [19], [37]. In addition, prior research has emphasized the significant role of third-party analytics and advertising libraries in shaping app privacy behavior. Studies such as Han et al. in [19] and Derr et al. in [32] showed that embedded SDKs can introduce additional data collection mechanisms that are not always clearly reflected in developer disclosures. As a result, privacy labels may capture potential data practices associated with integrated components rather than directly observable behavior within the app package. These findings suggest that, while Data Safety labels represent a promising step toward improving transparency, they may still face challenges similar to those identified in earlier studies on privacy policies and mobile app disclosures.

*Positioning of our Study:* While prior work has extensively examined mobile privacy from perspectives such as permissions, data flows, third-party tracking, and privacy policies, relatively few studies have focused on the direct comparison between marketplace transparency mechanisms and observable application-level indicators. Existing studies on Data Safety labels remain limited and often do not integrate systematic static analysis with structured evaluation of disclosure consistency. This study addresses this gap by providing a configuration-aware and indicator-based comparison between developer-reported Data Safety disclosures and static privacy indicators extracted from APK files. In addition, it extends prior work by analyzing both children-oriented and general-audience apps, and by applying statistical analysis techniques to evaluate the significance of observed differences. By combining insights from software analysis and marketplace transparency research, this work contributes a more comprehensive understanding of the reliability of privacy disclosures in modern mobile ecosystems.

## III. RESEARCH METHODOLOGY

This study adopts an application-centric static analysis approach to examine the consistency between Google Play Data Safety disclosures and observable technical indicators within Android apps. The methodology follows a structured empirical workflow consisting of four main phases: (1) design the study protocol, (2) APK collection and static inspection, (3) Data Safety label collection from Google Play, and (4) data analysis. Figure 1 presents an overview of the research workflow, illustrating all the steps undertaken to conduct this study.

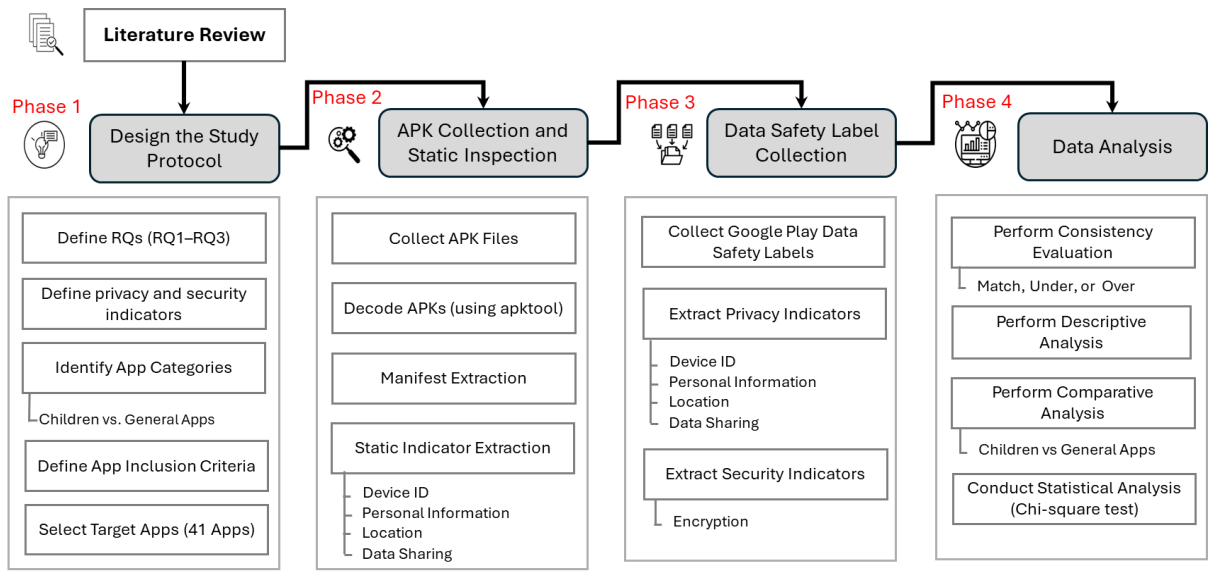

**FIGURE 1.** An Overview of the Research Workflow

### A. PHASE 1: DESIGNING THE STUDY PROTOCOL

The first phase focuses on establishing a systematic protocol to guide data collection and analysis. The protocol was designed based on prior literature on mobile privacy analysis and platform transparency mechanisms. First, the RQs (RQ1–RQ3), presented in Section I, were formulated to define the scope and objectives of the study. Based on these questions, a set of observable privacy and security indicators was defined to guide the extraction and comparison process. This phase also involved the following:

#### 1) PRIVACY AND SECURITY INDICATORS

To systematically evaluate disclosure accuracy, the study defined a set of observable indicators derived from Android application artifacts and Data Safety disclosures. These indicators were informed by prior research on Android privacy and security analysis, particularly studies focusing on permission usage and data access behaviors [38], [39], [40], and [41]. The selected indicators include:
  1. Location data collection.

2. Device identifiers (Device ID).
    3. Personal information access.
    4. Data sharing practices.
    5. Encryption in transit (as declared in the Data Safety label)

These indicators were selected to enable a systematic comparison between static app characteristics and Google Play Data Safety disclosures. By focusing on these categories, the study provides a structured way to evaluate whether developer-reported privacy practices align with observable indicators within the app binaries, offering a more comprehensive perspective on privacy transparency beyond permission-based analysis alone.

2) IDENTIFYING APP CATEGORIES

To support comparative analysis, the selected apps were categorized into two groups: children-oriented apps and general-audience mobile games. This categorization was based on Google Play age ratings, store descriptions, and developer-provided metadata. The classification enables the investigation of potential differences in disclosure practices across app types.

3) APP INCLUSION CRITERIA

To enable a meaningful empirical evaluation, applications were selected based on a set of predefined inclusion criteria. Specifically, the selected apps must: (i) belong to the mobile gaming category in the Google Play marketplace, (ii) demonstrate high popularity based on installation counts, user ratings, and store rankings, ensuring that the dataset reflects widely used applications, (iii) be clearly classified as either children-oriented or general-audience, and (iv) have an accessible APK file that can be successfully analyzed using static analysis techniques.

4) APP SELECTION

Based on these criteria, a dataset of 41 Android mobile gaming apps was constructed, consisting of 21 children-oriented apps and 20 general-audience games. The selected apps represent a diverse set of genres and development frameworks within the mobile gaming ecosystem. To support responsible reporting and avoid potential bias, applications are referenced using anonymized identifiers (i.e., **App1-App41**), with the mapping between identifiers and actual app names provided in the Appendix.

*B. PHASE 2: APKS COLLECTION AND STATIC INSPECTION*

The second phase involved four main steps: 1) APK collection, 2) APK decoding, 3) manifest extraction, and 4) static indicator extraction. This phase involved the following:

1) APKS COLLECTION

APK files for the selected apps were obtained from publicly available sources to enable static inspection. In some cases, apps were distributed in split formats (e.g., XAPK or APKM), which required additional processing to reconstruct a usable APK file. Some downloaded packages were incomplete or corrupted, as indicated by identical file sizes or extraction errors. These issues were addressed by re-downloading the apps from alternative sources while ensuring consistency with the selected app versions. This step ensured that all analyzed APKs were valid and suitable for further inspection.

2) DECODE APKS

The collected APK files were decoded using static analysis tools (i.e., apktool) to enable access to their internal resources and configuration files. The decoding process extracted readable representations of app components, including the manifest file and resource directories. In some cases, decoding failures occurred due to obfuscation or packaging inconsistencies, requiring repeated attempts or the use of alternative tool configurations. Successfully decoded APKs were then used as the basis for extracting privacy-related attributes.

3) MANIFEST EXTRACTION

The AndroidManifest.xml file was extracted from each decoded APK as the primary source of application-level configuration and permission declarations. This file was systematically inspected to identify privacy-relevant attributes, including declared permissions, app components, and network-related configurations. Variations in manifest structure across apps occasionally required manual verification to ensure accurate interpretation. The extracted manifest data provided a consistent foundation for identifying observable privacy indicators across all apps.

4) STATIC INDICATOR EXTRACTION

Static privacy indicators were derived from the extracted manifest data by mapping specific permissions and configuration attributes to predefined categories, including device IDs, personal information, location access, and data sharing. The extraction process involved both automated parsing and manual validation to ensure accuracy and consistency across apps.

*C. Phase 3: Collection of Google Play Data Safety Labels*

This phase involved collecting Data Safety labels for each app, which are publicly available on the Google Play marketplace. The Google Play Data Safety section was manually inspected to record developer-declared privacy practices. Based on the study protocol, the following disclosure fields were extracted: (i) location data collection, (ii) device identifier collection, (iii) personal information collection, (iv) app activity tracking, and (v) encryption in transit. All extracted information was systematically recorded in a structured dataset to ensure consistency and enable reliable comparison across apps. The complete dataset is provided in the Appendix.

*D. Phase 4: DATA SYNTHESIS AND COMPARATIVE ANALYSIS*

The final phase of the methodology focuses on analyzing the extracted data to evaluate the consistency between static indicators and Data Safety disclosures. This phase includes 1) consistency evaluation, 2) descriptive analysis, 3) comparative analysis, and 4) statistical testing to identify patterns and assess the significance of observed discrepancies.

### 1) CONSISTENCY EVALUATION

All extracted attributes were organized into a structured dataset consisting of 41 app records and multiple privacy indicators. For each indicator, a systematic comparison was performed between the Google Play Data Safety disclosures and the corresponding static analysis evidence derived from APK inspection. To characterize the relationship between reported and observable practices, three comparison outcomes were defined as follows:

- **Match:** The static analysis evidence is consistent with the developer-reported Data Safety disclosure.
- **Under-disclosure:** The static analysis indicates the presence of data access or collection that is not declared in the Data Safety label.
- **Over-disclosure:** The Data Safety label declares data collection or usage that is not supported by observable static evidence.

This step forms the core of the analysis by establishing a consistent framework for identifying discrepancies between developer disclosures and observable application characteristics.

### 2) DESCRIPTIVE STATISTICS

Descriptive statistics were used to quantify the prevalence of the defined comparison outcomes across the dataset. The analysis focused on overall mismatch frequency, indicator-specific discrepancy rates, and distribution patterns across privacy categories. This step enables the identification of systematic inconsistencies between marketplace disclosures and observable application characteristics, providing an initial understanding of how disclosure accuracy varies across different types of data.

### 3) COMPARATIVE ANALYSIS

To further investigate potential differences across app types, the dataset was divided into two categories: children-oriented apps and general-audience mobile games. Comparative analysis was then performed to examine variation in mismatch rates across these categories for each privacy indicator. This step supports the evaluation of whether disclosure inconsistencies differ depending on app audience, directly addressing the study's comparative research objectives.

### 4) STATISTICAL ANALYSIS

To assess whether the observed differences between app categories are statistically significant, Chi-square tests of independence were performed using Microsoft Excel [42]. The test evaluates the relationship between categorical variables, specifically disclosure outcome (match vs. mismatch) and app category (children-oriented vs. general-audience). By applying this statistical method, the study moves beyond descriptive observations and provides a rigorous assessment of whether observed differences reflect meaningful patterns or occur due to random variation. This step strengthens the validity of the findings and supports more robust conclusions regarding the nature of disclosure inconsistencies.

*Ethical Considerations:* The study relies exclusively on static inspection of publicly distributed Android APKs. No runtime interaction with user accounts or personal data was performed. All results are reported in aggregate form, and apps are anonymized to minimize potential harm to developers or users.

## IV. FINDINGS

This section reports the observed security and privacy indicators across the analyzed mobile gaming apps. The results are presented to highlight patterns in data collection, data sharing, and configuration-related practices extracted through static analysis. In addition, the findings examine the level of consistency between these observable indicators and the corresponding Data Safety disclosures. This analysis provides an empirical basis for identifying potential mismatches and understanding their distribution across different privacy categories.

### A. CONSISTENCY BETWEEN STATIC INDICATORS AND DATA SAFETY LABELS

The comparison between static indicators and Data Safety disclosures revealed varying levels of consistency across privacy categories. As shown in Table 1, the highest level of consistency was observed for device ID disclosures, with an agreement rate of 87.8% and a mismatch rate of 12.2% (i.e., 7.3% for under-disclosure, 4.9% for over-disclosure). In contrast, lower levels of consistency were observed for the remaining indicators. Data sharing practices exhibited a mismatch rate of 36.6%, while personal information disclosures showed a higher mismatch rate of 46.3% (i.e., 29.3 for under-disclosure, and 17.0% for over-disclosure). The largest discrepancy was observed for location-related disclosures, where 56.1% of apps showed inconsistencies between static analysis results and the corresponding Data Safety labels. These findings indicate that the accuracy of privacy disclosures varies considerably depending on the type of data involved. The relatively high consistency observed for device identifiers may be attributed to the direct relationship between advertising identifiers and well-defined platform permissions, making them easier for developers to report accurately. In contrast, categories such as location, personal information, and data sharing often involve more complex interactions with third-party libraries and backend services, which may introduce ambiguity in how data practices are interpreted and disclosed. As a result, these categories are more prone to inconsistencies between developer-reported information and observable app characteristics.

**TABLE 1.** Summary of Disclosure Consistency Across Privacy Indicators

| Indicator | Match | Under-disclosure | Over-disclosure |
|---|---|---|---|
| Device ID | 87.8% | 7.3% | 4.9% |
| Personal Info | 53.7% | 29.3% | 17.0% |
| Location | 43.9% | 0% | 56.1% |
| Data Sharing | 63.4% | 36.6% | 0% |

To provide a more concrete illustration of these discrepancies, Table 2 presents a detailed comparison between static indicators and Data Safety labels for a subset of the analyzed apps (first 10 apps). This example demonstrates how mismatches occur at the individual app level, including cases of both under-disclosure and over-disclosure across different privacy indicators.

**TABLE 2.** Example Comparison Between Static Indicators and Data Safety Labels (First 10 Apps)

| App ID | Device ID (S/L) | Personal Info (S/L) | Location (S/L) | Data Sharing (S/L) |
|---|---|---|---|---|
| App1 | Yes / No | Yes / No | No / No | Yes / No |
| App2 | Yes / Yes | Yes / Yes | No / No | Yes / No |
| App3 | Yes / Yes | Yes / Yes | No / Yes | Yes / Yes |
| App4 | Yes / Yes | Yes / No | No / No | Yes / No |
| App5 | Yes / Yes | Yes / No | No / No | Yes / No |
| App6 | Yes / Yes | Yes / No | No / Yes | Yes / No |
| App7 | Yes / Yes | Yes / No | No / Yes | Yes / Yes |
| App8 | Yes / Yes | Yes / Yes | No / Yes | Yes / No |
| App9 | Yes / No | Yes / No | No / No | Yes / No |
| App10 | Yes / Yes | Yes / Yes | No / No | Yes / No |

S denotes static analysis results, while L denotes developer-reported Data Safety disclosures.

### B. COMPARISON BETWEEN CHILDREN AND GENERAL GAMES

To better understand whether disclosure inconsistencies differ across application types, the dataset was divided into two categories: children-oriented apps and general-audience mobile games. The mismatch rates for each privacy indicator across the two categories are summarized in Table 3. The results indicate that the relationship between app category and disclosure inconsistency varies across privacy indicators rather than following a uniform pattern. As shown in Table 3, the mismatch rate for device identifier disclosures was 19.0% for children-oriented apps, compared to 5.0% for general-audience apps. Similarly, data sharing disclosures showed mismatch rates of 42.9% for children-oriented apps and 30.0% for general-audience apps. In contrast, personal information disclosures exhibited a slightly higher mismatch rate among general applications (50.0%) compared to children-oriented apps (42.9%). A similar trend was observed for location-related disclosures, where general applications showed a higher mismatch rate (60.0%) than children-oriented apps (52.4%). These results indicate that while some indicators (e.g., device identifiers and data sharing) show higher mismatch rates in children-oriented applications, others (e.g., personal information and location) are more inconsistent in general applications. Overall, disclosure inconsistencies are observed across both categories, suggesting that such discrepancies are not confined to a specific type of application but rather reflect broader challenges in accurately reporting privacy practices within the Data Safety framework.

**TABLE 3.** Comparison of Mismatch Rates Between App Categories

| Indicator | Children App Mismatch | General App Mismatch |
|---|---|---|
| Device ID | 19.0% | 5.0% |
| Personal Info | 42.9% | 50.0% |
| Location | 52.4% | 60.0% |
| Data Sharing | 42.9% | 30.0% |

### C. STATISTICAL SIGNIFICANCE ANALYSIS

To determine whether the observed differences in disclosure mismatches between children-oriented and general applications are statistically significant, a Chi-square test of independence was performed for each privacy indicator. For each indicator, contingency tables were constructed comparing the number of matching and mismatching cases across the two app categories. The results of the statistical analysis are summarized in Table 4. As shown in Table 4, none of the examined indicators exhibits a statistically significant association between app category and disclosure mismatch. Specifically, device identifier mismatches yielded $\chi^2 = 0.804$ (p = 0.370), data sharing mismatches yielded $\chi^2 = 0.281$ (p = 0.596), personal information mismatches yielded $\chi^2 = 0.021$ (p = 0.885), and location mismatches yielded $\chi^2 = 0.031$ (p = 0.860). These findings suggest that disclosure inconsistencies are not dependent on application category, but rather reflect broader, systemic limitations in the accuracy of marketplace privacy disclosures.

**TABLE 4.** Chi-Square Test Results for Disclosure Consistency

| Indicator | $\chi^2$ Value | p-value |
|---|---|---|
| Device ID | 0.804 | 0.370 |
| Personal Info | 0.021 | 0.885 |
| Location | 0.031 | 0.860 |
| Data Sharing | 0.281 | 0.596 |

### D. SECURITY PRACTICES IN DATA SAFETY LABELS

In addition to privacy-related disclosures, the analysis also examined security-related information reported in the Data Safety labels, particularly whether user data is encrypted during transmission. The results indicate variability in how mobile gaming apps report their data transmission security practices. While a majority of apps declared that data is encrypted in transit, a small subset of applications explicitly reported that user data is not encrypted in transit, including App35, and App40, as in Figure 2. Although encryption is not directly observable through static analysis, this finding provides additional insight into how developers communicate security practices through the Data Safety framework. Encryption of data in transit is widely considered a fundamental security measure for protecting user data from interception and unauthorized access [43], [44]. The observed variability suggests that, similar to privacy disclosures, security-related declarations may also lack consistency, highlighting an additional dimension of transparency challenges within mobile app marketplaces.

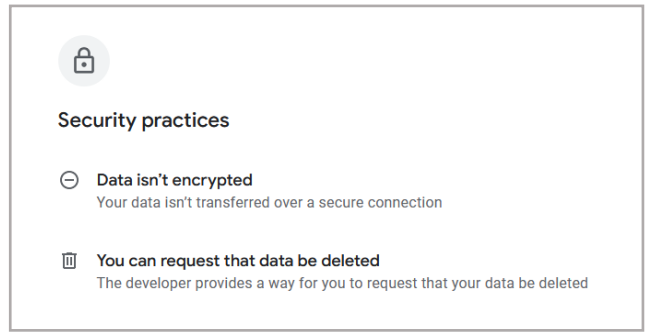

**FIGURE 2.** Examples of apps reporting that data is not encrypted in transit in Google Play Data Safety labels

## V. DISCUSSION

This section discusses the key findings of the study. It is important to note that the analysis is based on observable static indicators extracted from application artifacts rather than runtime behavior. Accordingly, the interpretations reflect potential privacy exposures inferred from static analysis rather than dynamic data flows.

### A. RELIABILITY OF DATA SAFETY LABELS

The results of this study indicate that the reliability of Google Play Data Safety disclosures varies across different types of privacy indicators. Device identifier disclosures demonstrated the highest level of consistency between developer-reported labels and observable static indicators, with an agreement rate of 87.8%. This relatively high agreement suggests that developers may find it easier to accurately report the use of device identifiers, as these practices are typically associated with well-defined permissions and standardized advertising frameworks. In contrast, other indicators exhibited substantially higher mismatch rates, particularly for location (56.1%), personal information (46.3%), and data sharing (36.6%). A closer examination of mismatch types reveals distinct patterns across indicators. Location-related discrepancies were primarily driven by over-disclosure, where developers report data collection that is not supported by observable static evidence. Conversely, data sharing inconsistencies were predominantly associated with under-disclosure, indicating that data exchange mechanisms may not always be fully reported in Data Safety labels. Personal information disclosures exhibited a combination of both under- and over-reporting. These findings suggest that the reliability of Data Safety disclosures is not uniform but depends strongly on the type of data being reported, reflecting differences in how developers interpret and implement reporting requirements.

These findings are consistent with prior research examining transparency mechanisms in mobile app ecosystems. Several studies have reported inconsistencies between developer-reported privacy policies and observable app behavior, particularly in relation to permission usage and data flows [10], [35]. Similarly, recent work analyzing Google Play Data Safety disclosures has suggested that developers may either over-report or under-report certain data practices due to ambiguity in the reporting guidelines or uncertainty about third-party library behavior [12]. The results of this study extend these observations by providing empirical evidence within the context of mobile gaming apps and by systematically comparing Data Safety disclosures with static privacy indicators extracted directly from APK files. This comparison highlights that while Data Safety labels represent an important step toward improving transparency, their reliability may still depend heavily on how developers interpret and implement the reporting requirements.

### B. PRIVACY TRANSPARENCY IN CHILDREN-ORIENTED APPS

An important objective of this study was to investigate whether privacy disclosure inconsistencies differ between children-oriented and general-audience apps. The results indicate that the relationship between app category and disclosure inconsistency is not uniform across privacy indicators. For certain indicators, such as device identifiers and data sharing, children-oriented apps exhibited higher mismatch rates. For example, device ID mismatches were observed in 19.0% of children-oriented apps compared to 5.0% of general-audience apps, while data sharing mismatches were 42.9% and 30.0%, respectively. In contrast, personal information and location-related disclosures showed higher mismatch rates among general-audience apps. Specifically, personal information mismatches were 50.0% for general apps compared to 42.9% for children-oriented apps, while location-related mismatches were 60.0% and 52.4%, respectively. These results indicate that inconsistencies in personal information and location disclosures are more pronounced in general-audience applications.

Despite these variations, statistical analysis confirmed that none of these differences are statistically significant, indicating that disclosure inconsistencies are present across both categories. These findings suggest that privacy transparency challenges are not limited to a specific group of apps but instead reflect broader issues in how developers interpret and report data practices within the Data Safety framework. This observation is consistent with prior studies showing that both children-oriented and general-audience apps may incorporate complex data collection mechanisms, including third-party analytics and advertising libraries, which can lead to inconsistencies between declared and actual practices [19], [45].

*C. SYSTEMIC NATURE OF DISCLOSURE INCONSISTENCIES*

The absence of statistically significant differences between children-oriented and general-audience apps suggests that inconsistencies between developer disclosures and observable indicators may represent a broader systemic issue within the mobile app ecosystem. Rather than being associated with a specific app category, these inconsistencies appear to arise from structural characteristics of modern mobile software development. This observation is consistent with recent empirical work [14], which shows that discrepancies between privacy disclosures and actual application behavior are widespread across mobile ecosystems. The study reports that numerous apps violate their declared privacy labels by transmitting undeclared data, indicating that such inconsistencies are systemic rather than isolated cases. This interpretation is further supported by prior work [13], which highlights that developers often face challenges in accurately completing privacy labels due to the complexity of modern mobile apps and limited visibility into third-party components. In particular, developers may struggle to understand and correctly report data collection behaviors, and tool-supported approaches based on static analysis have been proposed to improve label accuracy. In practice, many mobile apps rely heavily on third-party components such as advertising SDKs, analytics libraries, and cloud-based services. These components may introduce data collection capabilities that are not always clearly reflected in app permissions or static artifacts. Consequently, developers may report potential data practices associated with these components even when the corresponding behaviors are not directly observable within the application package. Taken together, these factors help explain the presence of indicator-specific mismatch patterns observed in this study, suggesting that disclosure inconsistencies are influenced not only by application category but also by the nature and complexity of the data being reported.

This interpretation aligns with previous research highlighting the complex role of third-party libraries in mobile privacy risks. Studies have shown that advertising and analytics frameworks frequently introduce additional data flows and tracking mechanisms that developers may not fully control or understand [32], [19]. As a result, accurately describing all potential data practices in marketplace disclosures can be challenging. Furthermore, platform reporting guidelines may encourage developers to disclose possible data usage scenarios rather than strictly observable behaviors, which may contribute to the mismatch patterns observed in this study. The findings therefore suggest that improving the accuracy of marketplace transparency mechanisms may require stronger tool support or automated validation approaches capable of identifying discrepancies between developer disclosures and observable app characteristics.

*D. SECURITY IMPLICATIONS OF DATA SAFETY DISCLOSURES*

In addition to privacy-related indicators, the analysis also examined security-related information reported in the Data Safety labels, particularly whether user data is encrypted during transmission. The results revealed variability in how mobile gaming apps report their encryption practices. While many apps indicated that user data is encrypted during transmission, several apps reported that user data may not be encrypted in transit. Encryption of data in transit is widely considered a fundamental security practice for protecting sensitive information from interception or unauthorized access. The presence of apps that report the absence of encryption mechanisms therefore raises potential concerns regarding the protection of user data in mobile gaming environments.

Previous studies examining security practices in mobile apps have similarly reported inconsistent adoption of secure communication mechanisms, particularly among apps that rely on third-party advertising or analytics services [44], [23]. In some cases, developers may depend on external services to manage data transmission security, which may lead to variations in how encryption practices are implemented or reported. The variability observed in this study suggests that Data Safety disclosures may provide useful signals regarding security practices but may not always present a complete picture of app security. From a user perspective, inconsistent reporting of encryption practices may make it difficult to assess the level of protection provided by different apps. These findings highlight the importance of continued research into automated methods for evaluating both privacy and security transparency in mobile app marketplaces.

*E. IMPLICATIONS FOR MOBILE PRIVACY TRANSPARENCY*

The findings of this study have several implications for improving transparency mechanisms in mobile app marketplaces. First, the observed inconsistencies between static indicators and Data Safety disclosures suggest that developers may face challenges in accurately interpreting the reporting requirements introduced by Google Play. Recent study has shown that even when privacy labels are available, they may not fully address users' information needs. Furthermore, the work demonstrated that many user privacy questions remain unanswered by current label designs, highlighting limitations in their ability to support informed decision-making [9]. In particular, the higher mismatch rates observed for indicators such as location access, personal information, and data sharing highlight the difficulty of mapping app behavior and third-party library functionality to the categories defined in the Data Safety framework. Similar challenges have been identified in prior studies examining the relationship between app permissions, privacy policies, and developer disclosures [39], [6]. These results suggest that additional guidance or automated tools may be required to assist developers in generating accurate Data Safety reports. For example, developer tools that automatically analyze app artifacts and recommend disclosure categories could reduce ambiguity and improve the reliability of privacy reporting. In particular, the distinct mismatch patterns observed across different indicators (e.g., over-disclosure in location and under-disclosure in data sharing) suggest that reporting challenges may be closely tied to how specific data categories are defined and interpreted within the Data Safety framework.

The results also have implications for platform providers and researchers investigating mobile privacy transparency. For platform providers such as Google Play, the mismatch patterns identified in this study suggest that self-reported privacy disclosures alone may not fully capture app behavior. Integrating automated verification mechanisms into the marketplace submission process could help identify inconsistencies between declared data practices and observable app characteristics. For example, automated checks based on static or dynamic analysis could flag potential mismatches during app submission and prompt developers to review their disclosures. From a research perspective, the findings highlight the importance of combining multiple analysis approaches, including static analysis, dynamic analysis, and marketplace metadata analysis, in order to better understand privacy transparency in mobile ecosystems. Future research may extend this work by developing automated techniques for identifying discrepancies between app behavior and marketplace disclosures at larger scale.

## VI. THREATS TO VALIDITY

As with any empirical study involving software artifacts and marketplace disclosures, several factors may influence the interpretation of the results. Following established guidelines for empirical software engineering research, threats to validity are categorized into internal, construct, and external validity.

### A. INTERNAL VALIDITY

Internal validity concerns whether the observed results may be influenced by methodological limitations during data collection and analysis. In this study, privacy indicators were extracted through static analysis of APK files, which may not capture behaviors triggered dynamically at runtime. For example, certain privacy-relevant actions may occur through remote services or dynamically loaded components that are not visible within the app package. Similar limitations of static analysis approaches have been reported in prior Android security studies [10], [11]. Another potential threat relates to APK availability and format variations. Some apps were initially obtained in formats such as XAPK or APKM, requiring additional extraction steps before analysis. In several cases, incomplete packages or identical file sizes indicated invalid downloads, which required replacing a small number of apps while maintaining the dataset balance. To mitigate these risks, permissions were manually verified when automated extraction produced unexpected results.

### B. CONSTRUCT VALIDITY

Construct validity concerns whether the selected indicators accurately represent the privacy practices being studied. In this work, privacy behavior was approximated using static indicators derived primarily from Android permissions and observable APK artifacts. While these indicators provide insights into potential data access capabilities, they may not fully capture all privacy-related behaviors within mobile apps. Some forms of data collection may occur through embedded third-party libraries, analytics frameworks, or advertising SDKs that do not require explicit permissions. As a result, certain privacy practices may remain undetected through permission-based analysis. Additionally, the Google Play Data Safety labels represent developer-reported disclosures rather than independently verified measurements of app behavior. Developers may declare potential data practices associated with integrated libraries even when those behaviors are not observable in the APK. Similar challenges in interpreting developer privacy disclosures have been reported in previous research on mobile privacy transparency [23], [27].

### C. EXTERNAL VALIDITY

External validity concerns the generalizability of the study findings beyond the analyzed dataset. This study analyzed 41 mobile gaming apps, including both children-oriented and general-audience games. Although this sample size is comparable to many empirical studies in mobile privacy research, it represents only a subset of the large Android app ecosystem. Another limitation relates to the app category scope. The study focuses specifically on mobile games, which may exhibit different privacy practices compared to other categories such as social networking, productivity, or financial apps. In addition, the results reflect a snapshot of apps and Data Safety disclosures at the time of data collection. Because mobile apps and marketplace disclosures evolve over time, future updates may lead to different observations. Despite these limitations, the dataset includes widely used apps with millions of downloads, providing a meaningful snapshot of privacy disclosure practices within the mobile gaming ecosystem [19]. Future studies involving larger datasets, additional app categories, and dynamic analysis techniques may further validate and extend the findings presented in this work.

## VII. CONCLUSIONS AND FUTURE WORK

Mobile apps increasingly rely on complex data collection, analytics, and advertising infrastructures to support functionality and monetization. To improve transparency regarding these practices, the Google Play marketplace introduced the Data Safety section, requiring developers to disclose how apps collect, share, and protect user data. Although this initiative represents an important step toward improving privacy transparency, the reliability of developer-reported disclosures remains largely dependent on self-reporting and may not always reflect observable app behavior. In this study, we conducted an empirical comparison between developer-reported Data Safety disclosures and static privacy indicators extracted from APKs in order to assess the consistency between marketplace transparency mechanisms and observable app characteristics. The analysis focused on a dataset of mobile gaming apps, including both children-oriented and general-audience games, enabling a comparative evaluation of privacy disclosure practices across app categories.

The study was conducted through a structured empirical process that involved extracting privacy-relevant indicators from APK files through static analysis and comparing these indicators with the corresponding disclosures reported in the Google Play Data Safety section. By examining indicators related to device identifiers, personal information access, data sharing practices, and location access, this work provides a systematic assessment of how accurately developer disclosures reflect observable privacy-related artifacts within mobile apps. The key findings of this study can be summarized as follows:

- Device identifier disclosures exhibited the highest level of agreement between static indicators and Data Safety labels, suggesting that some data practices may be easier for developers to report accurately due to their clear association with specific platform permissions and advertising frameworks.
- Other privacy indicators showed substantial inconsistencies, particularly for personal information, data sharing, and location-related disclosures, indicating that developer-reported transparency mechanisms do not always align with observable app artifacts.
- Children-oriented and general-audience apps exhibited similar patterns of disclosure inconsistencies, and statistical analysis revealed no significant differences between these categories, suggesting that transparency challenges extend across the broader mobile gaming ecosystem.
- Security-related disclosures, particularly regarding encryption practices, varied across apps, highlighting additional concerns regarding the clarity and reliability of information presented to users through the Data Safety framework.

These findings suggest that while the Google Play Data Safety section represents an important advancement in improving transparency in mobile app marketplaces, its effectiveness may be limited by the reliance on developer self-reporting and the inherent complexity of modern mobile app architectures. In particular, the widespread use of third-party analytics, advertising, and attribution frameworks may make it difficult for developers to precisely map app behavior to the categories defined within the Data Safety reporting framework. As a result, developers may disclose potential data practices associated with integrated libraries even when those practices are not directly observable in the app package.

Based on the empirical evidence, this study highlights several practical implications. First, developers may benefit from improved tool support and automated analysis mechanisms that assist in generating accurate Data Safety disclosures based on observable app artifacts. Second, platform providers such as Google Play may consider integrating automated validation mechanisms into the app submission process in order to detect inconsistencies between declared privacy practices and observable app characteristics. Such mechanisms could improve the reliability of marketplace transparency features and provide users with more accurate information when evaluating mobile apps. We believe this study opens several avenues for future research as follows:

1) The static analysis presented in this work can be extended with dynamic analysis techniques to observe runtime behavior, network communication patterns, and actual data flows. Combining static and dynamic analysis would enable researchers to evaluate whether the inconsistencies observed between Data Safety disclosures and static indicators correspond to real privacy risks during app execution.

2) Future studies could conduct large-scale measurements across hundreds or thousands of apps in order to evaluate the reliability of Data Safety disclosures across different app categories beyond mobile games, such as social networking, productivity, or financial apps.

3) Longitudinal studies could examine how developer disclosures evolve over time as apps are updated, monetization models change, and platform reporting requirements continue to evolve. Such studies would provide valuable insights into the stability and long-term effectiveness of marketplace transparency mechanisms.

4) Building on prior human-centered research, future work could explore how users interpret and rely on Data Safety labels when making installation decisions, and whether these disclosures meaningfully influence user perceptions of app privacy and security.

The results of this study can benefit multiple stakeholders. Researchers may use the findings to further investigate transparency mechanisms in mobile ecosystems and develop automated methods for detecting inconsistencies between app behavior and marketplace disclosures. Developers and app publishers may leverage the insights to improve privacy engineering practices and ensure that marketplace disclosures accurately reflect app behavior. Ultimately, this work contributes to a growing body of research indicating that improving transparency in mobile ecosystems requires not only clearer disclosure frameworks, but also stronger technical mechanisms for validating the information presented to users. In addition to the empirical findings presented in this study, this work contributes to the growing body of research investigating privacy transparency in mobile app ecosystems by providing a systematic comparison between developer-reported marketplace disclosures and observable app characteristics. While previous studies have examined privacy policies, permission usage, and third-party tracking infrastructures in mobile apps, fewer studies have focused specifically on the consistency between marketplace transparency mechanisms and technical indicators extracted from app artifacts. By combining static analysis of APK files with structured analysis of Google Play Data Safety disclosures, this study provides a novel perspective on the reliability of developer-reported privacy information. The insights generated through this comparison highlight the importance of strengthening transparency mechanisms in mobile marketplaces and demonstrate the value of combining software analysis techniques with marketplace metadata analysis in future privacy and security research.

## APPENDIX A

The list of mobile gaming apps used in our study with the full results for each app can, and the extended version of Table 2 be found in the following links: https://shorturl.at/mS5tW

## DECLARATION OF COMPETING INTEREST

The author declares that he has no known competing financial interests or personal relationships that could have appeared to influence the work reported in this article.

## ACKNOWLEDGMENTS

OpenAI's GPT-4 was minimally used for proofreading and grammar refinement during manuscript preparation.